\documentclass[runningheads]{llncs}
\usepackage[T1]{fontenc}
\usepackage{graphicx,verbatim}
\usepackage{amsmath}
\usepackage{amssymb} 
\usepackage{pifont} 
\newcommand{\xmark}{\ding{55}}
\begin{document}
\title{STA-Risk: A Deep Dive of Spatio-Temporal Asymmetries for Breast Cancer Risk Prediction}
%

\author{Zhengbo Zhou$^{\star}$  \qquad  Dooman Arefan$^{\dagger}$  \qquad  Margarita Zuley$^{\dagger}$  \qquad        \\     \qquad  Jules Sumkin$^{\dagger}$ \qquad Shandong Wu$^{\star \dagger \mathsection}$}  

\institute{$^{\star}$Intelligent Systems Program, University of Pittsburgh, Pittsburgh, PA, USA\\
    $^{\dagger}$Department of Radiology, University of Pittsburgh, Pittsburgh, PA, USA\\
    $^{\mathsection}$Department of Biomedical Informatics and Department of Bioengineering\\University of Pittsburgh, Pittsburgh, PA, USA}

\maketitle              
\begin{abstract}
Predicting the risk of developing breast cancer is an important clinical tool to guide early intervention and tailoring personalized screening strategies. Early risk models have limited performance and recently machine learning-based analysis of mammogram images showed encouraging risk prediction effects. These models however are limited to the use of a single exam or tend to overlook nuanced breast tissue evolvement in spatial and temporal details of longitudinal imaging exams that are indicative of breast cancer risk. In this paper, we propose STA-Risk (Spatial and Temporal Asymmetry-based Risk Prediction), a novel Transformer-based model that captures fine-grained mammographic imaging evolution simultaneously from bilateral and longitudinal asymmetries for breast cancer risk prediction. STA-Risk is innovative by the side encoding and temporal encoding to learn spatial-temporal asymmetries, regulated by a customized asymmetry loss. We performed extensive experiments with two independent mammogram datasets and achieved superior performance than four representative SOTA models for 1- to 5-year future risk prediction. Source codes will be released upon publishing of the paper.


\end{abstract}

\section{Introduction}

Breast cancer remains the most common cancer among women worldwide \cite{siegel2023cancer}. Early detection of breast cancer can significantly improve treatment outcomes and patient survival. In the setting of mammographic screening, assessing an individual’s risk of developing breast cancer is vital for early intervention and tailoring personalized screening strategies (e.g., risk-stratified screening frequency and/or supplemental screening). Early risk models such as Gail model \cite{quillin2004gail}, Claus model \cite{evans2007breast}, and Tyrer-Cuzick model \cite{himes2016breast}, only use basic personal and clinical data, showing limited performance. Machine learning-based analysis of screening mammogram images, such as the well-known Mirai model \cite{yala2021toward}, has shown superior performance than those early models. Mirai model is however, limited to the use of a single time-point mammogram exam as input, while most of the women in screening have multiple longitudinal exams.

Taking advantage of longitudinal screening mammograms for breast cancer risk prediction has a significant clinical value, as seen in several recent risk models. The PRIME+ \cite{lee2023enhancing} model used prior mammograms with a Transformer decoder, outperforming several methods that use only a single time-point exam. The LRP-NET \cite{dadsetan2022deep} followed developing asymmetry to integrate information from longitudinal mammogram exams and also showed increased performance than using a single mammogram exam. Karaman et al. \cite{karaman2024longitudinal} extended the Mirai model \cite{yala2021toward} to process multi-year mammogram exams, demonstrating that incorporating temporal trends can yield a higher prediction performance. While the ideas of using multiple longitudinal exams aim to capture breast tissue evolvement over time, it remains under-developed for effective methods that can capture fine-grained nuances of breast tissue variations in both the spatial and temporal dimensions to assess patient-specific breast cancer risk.

Bilateral asymmetry in mammogram images has been clinically recognized as a biomarker of risk for developing breast cancer, as subtle left–right breast differences often appear before overt lesions emerge \cite{scutt2006breast}. Early work by Zheng et al. \cite{zheng2012bilateral} \cite{zheng2014association} illustrated that simple measures of contralateral breast differences—such as mammographic density or pixel-level fluctuation—is associated with short-term breast cancer risk.  More recently, CNN- and RNN-based approaches have been developed for refined asymmetry modeling, sometimes integrating multi-year exams \cite{donnelly2024asymmirai} \cite{dadsetan2022deep}. RADIFUSION \cite{yeoh2023radifusion} extended similar ideas via specialized attention blocks and multi-view gating to capture longitudinal signals. BilAD \cite{shimokawa2023deep} used paired breast tomosynthesis images from both breasts to detect location-specific tissue differences. However, previous work on quantifying bilateral asymmetry is limited to oversimplified left–right breast comparisons, such as arithmetic subtraction or left–right feature pooling, which tend to overlook or preclude the potentially rich and predictive information in spatial and temporal imaging details. Moreover, these studies are mostly only applicable to a single time-point exam or an ordinary aggregation of multiple exams.

In evaluating longitudinal mammogram exams, subtle but clinically relevant tissue asymmetry patterns may emerge and evolve gradually, which requires a systematic mechanism to be capable of tracking in images both how each breast’s tissue changes on its own and how these changes constitute asymmetry with respect to the contralateral breast and over multiple sequential exams. In this paper, we propose STA-Risk (Spatial and Temporal Asymmetry-based Risk Prediction), a novel Transformer-based model that simultaneously captures both spatial and temporal asymmetries for breast cancer risk prediction. We aim to harness the bilateral (deviation between two breasts) and longitudinal (variations over time) cues to capture more expressive and fine-grained representation of mammographic imaging evolution, and associate them to the risk of developing breast cancer. The main contribution of this work is summarized as: \begin{itemize} \item 1)	We proposed a novel model structure, STA-Risk, that can systematically capture more expressive and fine-grained representation of mammographic imaging evolution at the bilateral and longitudinal dimensions for breast cancer risk prediction. \item 2)	We contributed a unified method that integrates side encoding and temporal encoding to learn spatial-temporal asymmetries, regulated by a customized asymmetry loss.
 \item 3)	We performed extensive experiments with two independent mammogram datasets and achieved superior performance than four representative SOTA models for 1- to 5-year future risk prediction.
 \end{itemize}

\begin{figure}[htbp]
    \centering
    \includegraphics[width=1\textwidth]{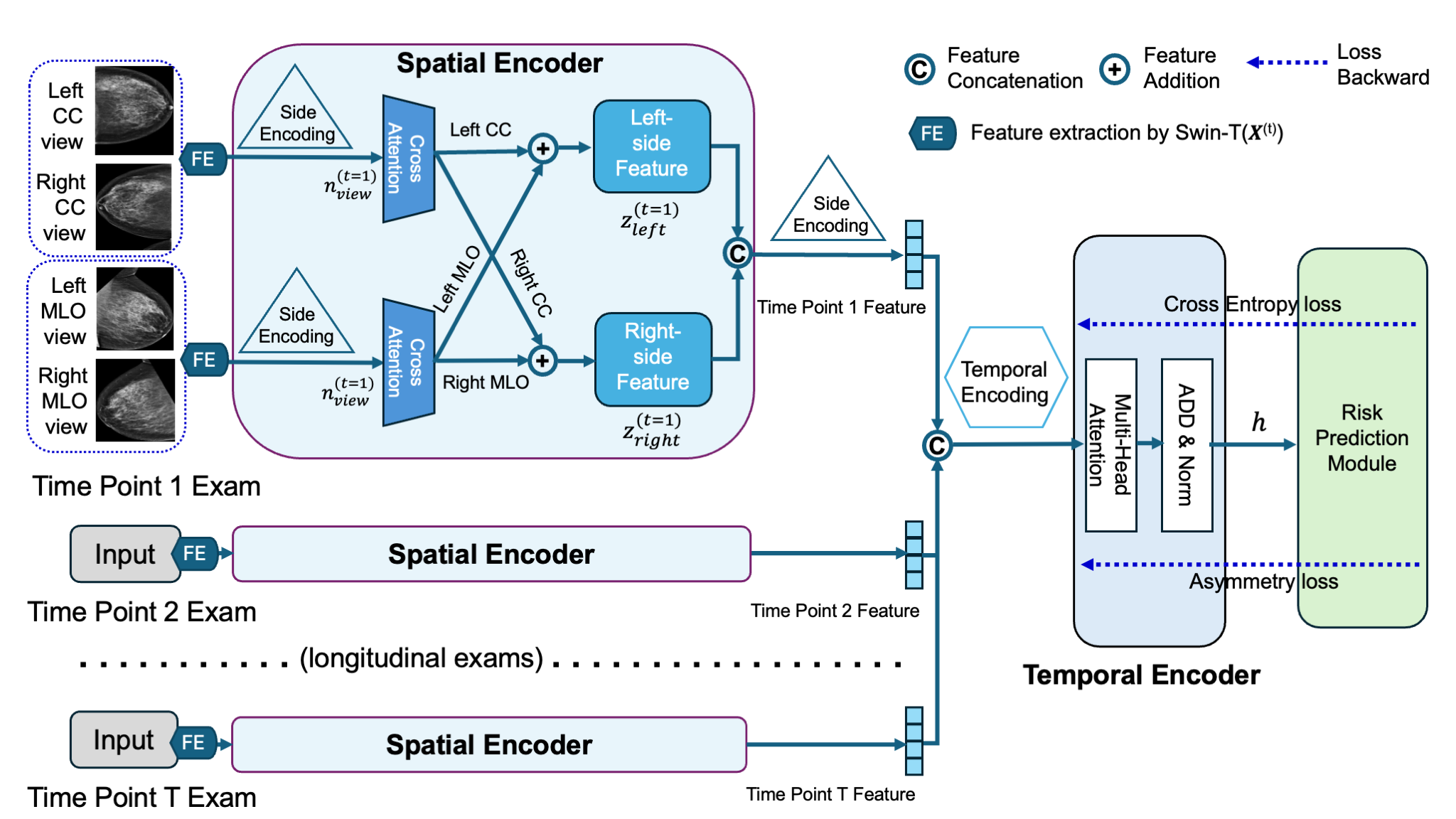}
    \caption{The proposed STA-Risk architecture that aims to capture spatial and temporal asymmetry from longitudinal screening mammogram exams for predicting breast cancer risk. The key components include side-aware spatial encoding, temporal attention, and a customized asymmetry loss.}
    \label{fig:example_figure}
\end{figure}

\section{Method}
The STA-Risk model (Figure 1) is characterized with three key components: \\
\textbf{Side encoding (left vs. right breast).} A learnable side-identifying vector is embeded into each patch token, preserving side-specific features and allowing the model to attend to side-preserved asymmetric cues.\\
\textbf{Temporal encoding.} Over multiple longitudinal exams, temporal encoding is designed to capture the evolving tissue patterns in each breast, taking into account when and how changes occur.\\
\textbf{Asymmetry loss.} A novel loss function is customized to selectively regularize left–right differences, preventing the model from discarding potentially valuable shared representations while guiding it to focus on clinically relevant asymmetries.

\subsection{Overall Network Architecture}
In our setting each patient has longitudinal mammogram exams imaged at $T$ consecutive time-points (years), with each exam consisting of four standard mammographic images/views: left cranio-caudal (LCC), left mediolateral oblique (LMLO), right cranio-caudal (RCC), and right mediolateral oblique (RMLO). As illustrated in Figure 1, the STA-Risk pipeline comprises a Spatial Encoder and a Temporal Encoder. At each time point, a side encoding is used to differentiate left from right breast. These inputs are then processed by the Spatial Encoder, which employs the attention mechanisms to extract relevant features and capture cross-side asymmetry. The resulting outputs are aggregated into a single-time-point feature vector that encapsulates the imaging patterns of each breast for an individual exam. Next, the multiple single-time-point feature vectors are fed into the Temporal Encoder, which relies on temporal encoding and side encoding to maintain the chronological imaging information and retains the breast-side distinction for consistent cross-breast comparisons. By learning how local imaging structures of breast tissue evolve over time and across left-right breast, the model detects progressive spatiotemporal changes and lateral discrepancies that are potentially related to breast cancer risk. The final output is then passed to a Risk Prediction Module, supervised by two losses: a cross-entropy loss that refines cancer-vs-normal status discrimination and a customized asymmetry loss that drives the network to exploit left–right and temporal breast tissue morphological differences. This prediction module utilizes an additive hazard \cite{yala2021toward} \cite{karaman2024longitudinal} and it integrates the complete history embedding, denoted as $h$, to estimate future risk of developing breast cancer. The cumulative risk over \( k \) years (where \( k \in \{1, 2, \ldots, 5\}\), as we target to predict 1- to 5-year risk) is computed by summing the baseline risk \( \beta_0 \) with the annual hazard terms \( \beta_1, \beta_2, \ldots, \beta_k \), where each \( \beta_j \) corresponds to the incremental risk for the \( j \)-th future year.

\subsection{Side Encoding}
\label{sec:side-encoding}

\noindent
A key of our method is to reserve the breast side information to enable side consistent bilateral analysis between images of the left and right side. Here a very different operation is to not use direct subtraction of pixel intensities to measure the left-right differences, as it will preclude in the first place some of the side-specific imaging details that reflect subtle changes but matter for risk assessment. Instead, we introduce a learnable side embedding, denoted by $\mathbf{v}_{\mathrm{left}}$ or $\mathbf{v}_{\mathrm{right}}$, which are added directly to feature embeddings. Specifically, at each imaging time-point $t$, each view mammogram \( \mathbf{X}_{\text{view}}^{(t)} \) is processed using the tiny Swin Transformer \cite{liu2021swin} feature extractor (denoted by Swin-T) and a side embedding $\mathbf{v}_{\mathrm{side}}$ is incorporated into the extracted features, expressed as: 
\begin{equation}
 n_{\text{view}}^{(t)} = \text{Swin-T}(\mathbf{X}^{(t)}) + v_{\text{side}}.
\end{equation}
We incorporate side encoding at two key places in the pipeline. First, prior to the cross-attention step in the Spatial Encoder, we attach a side embedding to the feature embedding to preserve left–right breast identity during spatial processing. Second, before entering the Temporal Encoder, each exam-specific feature vector is again augmented with a side embedding, enabling the model to monitor how each breast evolves over time. This two-stage approach empowers the network to detect subtle asymmetries in both the single-exam (spatial) structure and the longitudinal (temporal) progression of breast tissue. Overall our novel approach offers two main advantages: 1) Preserving Side-Specific Features. By encoding side identity with a separate embedding, our model retains a fine-grained feature representation of left- or right-specific imaging traits and analyzes them in a nonlinear manner. 2) Contextual Modeling of Asymmetry. The Transformer learns nuanced relationships, where a small local change in one breast may be highly predictive of abnormality, while bilaterally symmetric changes may be actually less alarming. Notably, this level of contextual nuances cannot be captured by simple arithmetic subtraction of left-right breasts.

\subsection{Temporal Encoding}
Sequential screening mammogram exams may occur at irregular intervals (e.g., 0.5, 1, 2, or even 3 years) due to irregular hospital visits, making purely index-based encodings suboptimal.
Inspired by \cite{sriram2021covid} and \cite{holste2024harnessing}, we capture irregular intervals between exams, we encode temporal information based on the time difference relative to the “present” exam at a reference-point (denoted by $y_p$). In our formulation, exam dates are converted from years to months. For the $t$-th exam, let $y_t$ denote its year, the relative time is calculated as \( \tau_t = 12 \times (y_t - y_p) \). This way for the present exam, \(\tau_t=0\), and for prior exams, \(\tau_t\) will be negative, representing the number of months prior to the current exam. We then derive a d-dimensional temporal embedding, denoted as \(\mathrm{TEmb}(\tau_t)\), by encoding each temporal input using alternating sine and cosine functions over the embedding dimensions \( i \in [0, \frac{d}{2}] \), where \( d \) is the total dimensionality of the temporal embedding vector:

\begin{equation}
\begin{aligned}
\mathrm{TEmb}(\tau_t)_{2i} &= \sin\!\Bigl(\frac{\tau_t}{10000^{\frac{2i}{d}}}\Bigr),\\[1ex]
\mathrm{TEmb}(\tau_t)_{2i+1} &= \cos\!\Bigl(\frac{\tau_t}{10000^{\frac{2i}{d}}}\Bigr),
\end{aligned}
\quad i = 0, 1, \ldots, \frac{d}{2}-1.
\end{equation}
This formulation ensures that exams that are temporally closer to the present exam have similar embeddings, while those further apart yield more distinct representations. The resulting temporal embedding is then added to the spatial and side encodings to form a unified representation for each exam.

\subsection{Asymmetry Loss}
Following the Spatial Encoder, each mammogram exam is reduced to two feature embeddings—one for the left and the other for the right breast—yielding
\(\{\mathbf{z}_{\text{left}}^{(t)}, \mathbf{z}_{\text{right}}^{(t)}\}\) 
for each of the \(T\) time points \(t \in \{1,2,3,...T\}\), following cross-attention between the corresponding $n^{(t)}_{\text{CC}}$ and $n^{(t)}_{\text{MLO}}$ views. These embeddings are fed into a Temporal Encoder. The final output at each time point encapsulates both local spatial information and initial left–right differences. To quantify cross-breast asymmetry, we measure the Euclidean distance between the left and right embeddings at each time point \(t\),
\begin{equation}
  D_t \;=\; \|\mathbf{z}_{\text{left}}^{(t)} - \mathbf{z}_{\text{right}}^{(t)}\|,
\end{equation}
and compute the average \(\overline{D}\) over \(T\) time points. In addition, we incorporate longitudinal (temporal) asymmetry by assessing how each breast side’s embedding changes between consecutive exams, for instance
\begin{equation}
  \Delta_t^{(\mathrm{left})} \;=\; \|\mathbf{z}_{\text{left}}^{(t)} - \mathbf{z}_{\text{left}}^{(t+1)}\|
  \quad (\text{and similarly for the right side}).
\end{equation}
For cancer cases (\(y=1\)), we expect both larger cross-breast discrepancies and greater exam-to-exam changes, whereas normal cases (\(y=0\)) should exhibit smaller cross-breast differences and less significant temporal variation. We compute the average \(\overline{\Delta}\) over \(T\) time points. We employ a margin-based hinge loss to provide target-specific guidance on these distances:
\begin{equation}
\mathcal{L}_{\text{asym}} =
\begin{cases}
\max\bigl(0, m_1 - \overline{D}\bigr)
+ \max\bigl(0, m_1' - \overline{\Delta}\bigr), & \text{if } y=1  \\
\\[-1em]
\max\bigl(0, \overline{D} - m_2\bigr)
+ \max\bigl(0, \overline{\Delta} - m_2'\bigr), & \text{if } y=0 ,
\end{cases}
\end{equation}
 where \(y\in\{0,1\}\) indicates the status of cancer vs. normal, and \(m_1,m_2,m_1',m_2'\) are user-defined margins. Let \(\mathcal{L}_{\text{primary}}\) denote the main objective, we integrate them to customize a total loss as:
\begin{equation}
  \mathcal{L}_{\text{total}}
  \;=\;
  \mathcal{L}_{\text{primary}}
  \;+\;
  \lambda \,\mathcal{L}_{\text{asym}},
\end{equation}
where \(\mathcal{L}_{\text{asym}}\) combines both cross-breast and longitudinal terms, and \(\lambda\) balances the strength of asymmetry constraints. This joint formulation enables the STA-Risk model to leverage bilateral discrepancies and progressive temporal changes in the prediction of risk.

\section{Experiments}
\subsection{Study Cohorts and Datasets}
Our experiments used two independent patient cohorts and imaging datasets. The first is the Karolinska Case-Control \textbf{(CSAW-CC) Dataset} \cite{Strand2022}, which is a part of the Cohort of Screen-Aged Women (CSAW). The CSAW-CC dataset was specifically curated for developing breast imaging-based AI tools. It includes women aged 40–74 years old who underwent mammographic screening between 2008 and 2016 using Hologic imaging systems. To mitigate potential bias in the risk prediction due to early cancer signs or early-detectable cancers, patients diagnosed with breast cancer within six months following the “present” screening exam were excluded. Our analysis included subjects who have at least two sequential screening exams. The final CSAW-CC cohort consisted of 406 breast cancer cases (all biopsy-proven) and 6,053 normal controls, with inter-exam intervals ranging from 12 to 36 months. The second dataset (denoted as \textbf{Independent Dataset}) is a retrospectively collected case-control cohort at a different hospital, with individuals who participated in routine breast cancer screening from 2007 to 2014 also using Hologic systems. We have data use agreement for this not-publicly-available dataset. This cohort comprises 293 breast cancer cases (all biopsy-proven) and 297 normal controls (at least 1-year follow-up to ensure normal status). Each subject had at least two sequential screening mammogram exams, with inter-exam intervals ranging from 12 to 24 months.

\subsection{Implementation Details}

STA-Risk models were trained to predict 1- to 5-year breast cancer risk using sequential screening mammograms. For each mammogram exam of a patient’s data, it is treated as a reference time-point (Prior 0) and we then traced backward up to maximum three prior exams (Prior 1, Prior 2, Prior 3), with irregular intervals of 12–36 months between consecutive exams. Patient outcomes (e.g., cancer vs. normal status) were determined based on the next follow-up exam occurring after $k$ years since Prior 0, where $k$ corresponds to the prediction horizon (1–5 years). All dataset splits were rigorously performed at the patient level to prevent data leakage.\\

We employed patient-wise 5-fold cross-validation to evaluate the performance of the proposed STA-Risk model. In each fold, data is split into  training and testing set in an 80-20 radio. STA-Risk was benchmarked against several related and representative risk models,  including LRP-NET \cite{dadsetan2022deep}, Primte+ \cite{lee2023enhancing}, and LoMaR \cite{karaman2024longitudinal} that all work on longitudinal exams. We also compared to the Mirai \cite{yala2021toward} model, even though it only works on a single time-point exam; as it is a well-known model and here it serves as a baseline to show the benefits of using longitudinal data. To make the model learning focus on breast regions, as a preprocessing we used Libra \cite{keller2012estimation} \cite{keller2015preliminary} to segment breasts first. To mitigate class imbalance, we adopt the reweighted cross-entropy loss function.  The model was trained for 30 epochs with a batch size of 32, and the best checkpoint was selected via a grid search over learning rate of 5e‑5 and 1e‑5. All experiments were conducted on an NVIDIA TESLA A100 GPU, courtesy of our institution's computing resources. The parameter $\lambda$ for the asymmetric loss was set to 0.01, and the parameters $m1$, $m2$, $m1'$, $m2'$ were all set to 1, as determined empirically through experiments. Model performance was evaluated using C-index and the Area Under the ROC Curve (AUC), with the mean AUC and standard deviations computed over 5-fold cross-validation for predicting the 1- to 5-year risk. 
\begin{table}[h!]
\centering
\caption{Prediction Performance Comparisons of the Proposed STA-Risk Model to Other Models on two  Datasets.}
\label{tab:model_performance}
\scriptsize
\resizebox{\textwidth}{!}{%
\begin{tabular}{|l|c|c|c|c|c|c|}
\hline
\textbf{Model} & \textbf{C-index} & \textbf{1-year AUC} & \textbf{2-year AUC} & \textbf{3-year AUC} & \textbf{4-year AUC} & \textbf{5-year AUC} \\
\hline\hline
\multicolumn{7}{|c|}{\textbf{CSAW-CC Dataset}} \\
\hline
Mirai \cite{yala2021toward} & 0.687 $\pm$ 0.02 & 0.705 $\pm$ 0.03 & 0.680 $\pm$ 0.02 & 0.664 $\pm$ 0.03 & 0.664 $\pm$ 0.02 & 0.630 $\pm$ 0.01 \\
LRP-NET \cite{dadsetan2022deep} & 0.670 $\pm$ 0.02 & 0.692 $\pm$ 0.03 & 0.688 $\pm$ 0.02 & 0.654 $\pm$ 0.01 & 0.654 $\pm$ 0.01 & 0.636 $\pm$ 0.02 \\
LoMaR \cite{karaman2024longitudinal} & 0.696 $\pm$ 0.01 & 0.709 $\pm$ 0.02 & 0.708 $\pm$ 0.01 & 0.689 $\pm$ 0.02 & 0.686 $\pm$ 0.03 & 0.675 $\pm$ 0.03 \\
Prime+ \cite{lee2023enhancing} & 0.683 $\pm$ 0.02 & 0.698 $\pm$ 0.01 & 0.697 $\pm$ 0.01 & 0.685 $\pm$ 0.01 & 0.681 $\pm$ 0.01 & 0.672 $\pm$ 0.01 \\
\textbf{STA-Risk} & \textbf{0.722 $\pm$ 0.01} & \textbf{0.749 $\pm$ 0.02} & \textbf{0.744 $\pm$ 0.01} & \textbf{0.706 $\pm$ 0.02} & \textbf{0.704 $\pm$ 0.03} & \textbf{0.694 $\pm$ 0.02} \\
\hline\hline
\multicolumn{7}{|c|}{\textbf{Independent Dataset}} \\
\hline
Mirai \cite{yala2021toward} & 0.685 $\pm$ 0.02 & 0.685 $\pm$ 0.03 & 0.660 $\pm$ 0.03 & 0.670 $\pm$ 0.02 & 0.655 $\pm$ 0.03 & 0.653 $\pm$ 0.02 \\
LRP-NET \cite{dadsetan2022deep} & 0.676 $\pm$ 0.02 & 0.683 $\pm$ 0.03 & 0.665 $\pm$ 0.03 & 0.645 $\pm$ 0.03 & 0.640 $\pm$ 0.02 & 0.630 $\pm$ 0.03 \\
LoMaR \cite{karaman2024longitudinal} & 0.706 $\pm$ 0.02 & 0.717 $\pm$ 0.02 & 0.691 $\pm$ 0.02 & 0.661 $\pm$ 0.03 & 0.634 $\pm$ 0.02 & 0.620 $\pm$ 0.02 \\
Prime+ \cite{lee2023enhancing} & 0.699 $\pm$ 0.02 & 0.702 $\pm$ 0.01 & 0.679 $\pm$ 0.03 & 0.644 $\pm$ 0.03 & 0.618 $\pm$ 0.03 & 0.597 $\pm$ 0.02 \\
\textbf{STA-Risk} & \textbf{0.732 $\pm$ 0.02} & \textbf{0.744 $\pm$ 0.02} & \textbf{0.717 $\pm$ 0.03} & \textbf{0.684 $\pm$ 0.03} & \textbf{0.662 $\pm$ 0.03} & \textbf{0.654 $\pm$ 0.03} \\
\hline
\end{tabular}%
}
\end{table}

\begin{table}[h!]
\centering
\caption{Ablation Study Results on the STA-Risk Model Components (Side – Side Encoding; Tmp – Temporal Encoding; Asy – Asymmetry Loss)}
\label{tab:ablation_study}
\scriptsize
\resizebox{\textwidth}{!}{%
\begin{tabular}{|c|c|c|c|c|c|c|c|c|}
\hline
\textbf{Side} & \textbf{Tmp} & \textbf{Asy} & \textbf{C-index} & \textbf{1-year AUC} & \textbf{2-year AUC} & \textbf{3-year AUC} & \textbf{4-year AUC} & \textbf{5-year AUC} \\
\hline\hline
\multicolumn{9}{|c|}{\textbf{CSAW-CC Dataset}} \\
\hline
\xmark & \xmark & \xmark & 0.693 $\pm$ 0.01 & 0.706 $\pm$ 0.01 & 0.707 $\pm$ 0.01 & 0.684 $\pm$ 0.01 & 0.691 $\pm$ 0.01 & 0.671 $\pm$ 0.02 \\
\checkmark & \xmark & \xmark & 0.699 $\pm$ 0.01 & 0.714 $\pm$ 0.01 & 0.715 $\pm$ 0.02 & 0.696 $\pm$ 0.01 & 0.691 $\pm$ 0.02 & 0.679 $\pm$ 0.02 \\
\xmark & \xmark & \checkmark & 0.695 $\pm$ 0.01 & 0.712 $\pm$ 0.01 & 0.705 $\pm$ 0.01 & 0.684 $\pm$ 0.01 & 0.680 $\pm$ 0.01 & 0.675 $\pm$ 0.01 \\
\checkmark & \xmark & \checkmark & 0.704 $\pm$ 0.02 & 0.718 $\pm$ 0.01 & 0.715 $\pm$ 0.01 & 0.697 $\pm$ 0.01 & 0.693 $\pm$ 0.01 & 0.683 $\pm$ 0.01 \\
\checkmark & \checkmark & \xmark & 0.705 $\pm$ 0.01 & 0.726 $\pm$ 0.01 & 0.722 $\pm$ 0.01 & 0.684 $\pm$ 0.01 & 0.686 $\pm$ 0.02 & 0.674 $\pm$ 0.01 \\
\checkmark & \checkmark & \checkmark & \textbf{0.722 $\pm$ 0.01} & \textbf{0.749 $\pm$ 0.01} & \textbf{0.744 $\pm$ 0.01} & \textbf{0.706 $\pm$ 0.02} & \textbf{0.704 $\pm$ 0.02} & \textbf{0.694 $\pm$ 0.02} \\
\hline\hline
\multicolumn{9}{|c|}{\textbf{Independent Dataset}} \\
\hline
\xmark & \xmark & \xmark & 0.707 $\pm$ 0.02 & 0.713 $\pm$ 0.02 & 0.688 $\pm$ 0.02 & 0.653 $\pm$ 0.03 & 0.628 $\pm$ 0.02 & 0.614 $\pm$ 0.02 \\
\checkmark & \xmark & \xmark & 0.717 $\pm$ 0.02 & 0.729 $\pm$ 0.02 & 0.704 $\pm$ 0.02 & 0.665 $\pm$ 0.03 & 0.639 $\pm$ 0.03 & 0.622 $\pm$ 0.03 \\
\xmark & \xmark & \checkmark & 0.725 $\pm$ 0.02 & 0.730 $\pm$ 0.02 & 0.706 $\pm$ 0.04 & 0.675 $\pm$ 0.04 & 0.650 $\pm$ 0.03 & 0.632 $\pm$ 0.03 \\
\checkmark & \xmark & \checkmark & 0.728 $\pm$ 0.02 & 0.739 $\pm$ 0.03 & 0.711 $\pm$ 0.04 & 0.680 $\pm$ 0.04 & 0.655 $\pm$ 0.04 & 0.643 $\pm$ 0.04 \\
\checkmark & \checkmark & \xmark & 0.725 $\pm$ 0.02 & 0.730 $\pm$ 0.02 & 0.706 $\pm$ 0.04 & 0.675 $\pm$ 0.04 & 0.650 $\pm$ 0.03 & 0.632 $\pm$ 0.03 \\
\checkmark & \checkmark & \checkmark & \textbf{0.732 $\pm$ 0.02} & \textbf{0.744 $\pm$ 0.02} & \textbf{0.717 $\pm$ 0.03} & \textbf{0.684 $\pm$ 0.03} & \textbf{0.662 $\pm$ 0.03} & \textbf{0.654 $\pm$ 0.03} \\
\hline
\end{tabular}%
}
\end{table}

\subsection{Results}
Table 1 presents the performance comparisons of the proposed STA-Risk model on the two datasets. As can be seen, our model consistently outperformed the compared models in predicting breast cancer risk across multiple time horizons. On the CSAW-CC dataset, STA-Risk achieves the highest C-index (0.722 ± 0.01), surpassing Mirai (0.687 ± 0.02), LRP-NET (0.670 ± 0.02), Prime+ (0.683
± 0.02), and LoMaR (0.696 ± 0.01). In terms of AUC, STA-Risk shows higher performance for all the future risk predicted from 1 to 5 years. A similar out- performing pattern of our STA-Risk model is also observed on the Independent dataset. These results collectively verify the robustness and generalizability of our model across the two different datasets.\\
Table 2 shows the ablation effects of the key components of our model. As can be seen, the combination of the three components yielded the highest performance. Furthermore, excluding temporal attention significantly reduces longer-term predictability, highlighting the importance of capturing breast tissue’s longitudinal evolvement and feature dependencies. Similarly, excluding asymmetry loss or side encoding also lowers performance, suggesting the rele- vance of bilateral asymmetry for risk prediction. Figure 2 visualized Grad-Cam heatmaps of several subjects in the test set: in the left panel, STA-Risk-produced heatmaps capture the suspicious lesions (despite false positives) later diagnosed as cancer as annotated by radiologists; notably in the right panel, Subject 3’s highlighted imaging features over longitudinal exams demonstrates our model is able to attend to subtle tissue asymmetrical progression consistently focused on the temporally evolving regions of potential caner development.

\begin{figure}[t!]
    \centering
    \includegraphics[width=1\textwidth]{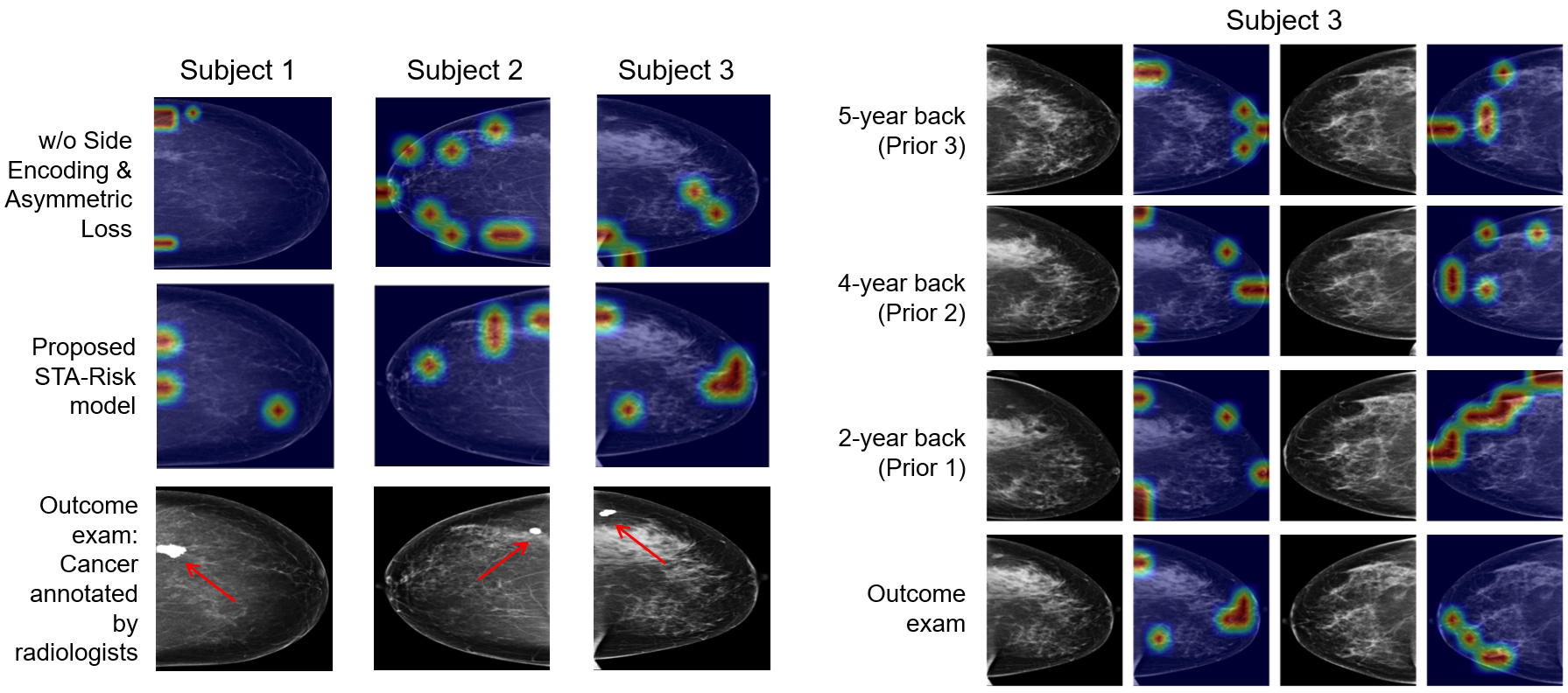}
    \caption{(Left) Visualization on the Differences when Using STA-Risk vs. Without Side Encoding and Asymmetric Loss. (Right) Illustrative Visualization on the Effects of the STA-Risk Model.}
    \label{fig:example_figure}
\end{figure}

\section{Conclusion}
In this work, we presented a novel model structure for asymmetry-aware breast cancer risk prediction using longitudinal screening mammograms. By combining temporal encoding to account for sequential data, and a learnable side encoder to retain nuance details between left–right breast, our attention-based model circumvents some of the limitations of previous models. Also, the asymmetric loss explicitly regularizes bilateral differences to adaptively learn risk-predictive imaging dissimilarities. Experimental results showed superior risk prediction performance that outperformed multiple SOTA models. Future work includes further evaluation of our model on larger and multi-center datasets. Notably, our model can account for a varying number of sequential exams for prediction the risk of from 1 to 5 years, which provides strong capacities for real-world uses while many patients have multiple screening exams often time not acquired with fixed time intervals. Our model can inform risk-stratified breast cancer screening to guide risk-reduction interventions and personalized screening strategies, towards detecting breast cancer early and saving lives.

\section{Acknowledgments}
\label{sec:acknowledgments}

This work was supported in part by a NIH Other Transaction research contract \#1OT2OD037972-01, the grant \#1R01EB032896 (and a Supplement grant \#3R01EB032896-03S1) as part of the NSF/NIH Smart Health and Biomedical Research in the Era of Artificial Intelligence and Advanced Data Science Program, a NSF grant (CICI: SIVD: \#2115082), an Amazon Machine Learning Research Award, and the University of Pittsburgh Momentum Funds (a scaling grant) for the Pittsburgh Center for AI Innovation in Medical Imaging. This work used Bridges-2 at Pittsburgh Supercomputing Center through allocation [MED200006] from the Advanced Cyberinfrastructure Coordination Ecosystem: Services \& Support (ACCESS) program, which is supported by NSF grants \#2138259, \#2138286, \#2138307, \#2137603, and \#2138296. This research was also supported in computing resources by the University of Pittsburgh Center for Research Computing and Data, RRID:SCR\_022735, through the resources provided by the H2P cluster, which is supported by NSF award number OAC-2117681. The views and conclusions contained in this document are those of the authors and should not be interpreted as representing official policies, either expressed or implied, of the NIH or NSF.

%
%
%
\bibliography{mybibliography}
\bibliographystyle{splncs04}

\end{document}